\documentclass{sig-alternate-10pt}
\usepackage{graphicx}
\usepackage[justification=centering]{caption}
\usepackage[format=hang]{subcaption}
\usepackage{url}
\usepackage{amsfonts}
\usepackage{times}
\usepackage{threeparttable}
\usepackage{amsmath, amssymb}
\usepackage{pslatex}
\usepackage{enumerate}
\usepackage{algorithm}
\usepackage{clrscode}
 \usepackage{url}
\usepackage{arcs}
\usepackage{yhmath}
\numberofauthors{2}

\usepackage{epstopdf}

\usepackage[toc,page]{appendix}
\usepackage{color} % enable color, not necessary

\begin{document}

%\conferenceinfo{SIGMETRICS'11,} {June 7--11, 2011, San Jose,
%California, USA.} \CopyrightYear{2011}
%\crdata{978-1-4503-0262-3/11/06} \clubpenalty=10000 \widowpenalty =
%10000

%\title{untitled }

%\author{
%Simon S. Lam and Chen Qian\\
%Department of Computer Science,
%The University of Texas at Austin \\
%Austin, Texas 78712 \\
%\{lam, cqian\}@cs.utexas.edu \\
%}
%\numberofauthors{3}
%\author{ Ye Yu, Qian Chen and Xin Li \\
%\affaddr{Department of Computer Science, University of Kentucky} \\
%\affaddr{ Austin, Texas, 78712} \\
%\email{Sheldon Cooper},
%\and
% Sheldon Cooper \\
%\affaddr{Department of Computer Science, The University of Texas at Austin} \\ \affaddr{ Austin, Texas, 78712} \\
%\email{Sheldon Cooper},
%\and
% Sheldon Cooper \\
%\affaddr{Department of Computer Science, The University of Texas at Austin} \\ \affaddr%{ Austin, Texas, 78712} \\
%\email{Sheldon Cooper}
 %}

\title{Distributed Collaborative Monitoring in Software Defined Networks}

\author{
Ye Yu \ \ \ \ \ \ \ \  \ \ \ \ \ \ \ \ \ \ \ Chen Qian\ \ \ \ \ \  \ \ \ \ \ \ \ \ \ \ \ \ \ \ Xin Li \\
ye.yu@uky.edu\ \ \ \ \ \ \ \ qian@cs.uky.edu \ \ \ \ \ \ \ \ \ xin.li@uky.edu\\
Department of Computer Science\\ University of Kentucky\\
%Lexington, Kentucky, 40506
}

\maketitle

\begin{abstract}
We propose a Distributed and Collaborative Monitoring system, DCM, with the following properties. First, DCM allows switches to collaboratively achieve flow monitoring tasks and balance measurement load. Second, DCM is able to perform per-flow monitoring, by which different groups of flows are monitored using different actions. Third, DCM is a memory-efficient solution for switch data plane and guarantees system scalability.   DCM uses a novel two-stage Bloom filters to represent monitoring rules using small memory space. It utilizes the centralized SDN control to install, update, and reconstruct the two-stage Bloom filters in the switch data plane. We study how DCM performs two  representative monitoring tasks, namely flow size counting and packet sampling, and evaluate its performance. Experiments using real data center and ISP traffic data on real network topologies show that DCM achieves highest measurement accuracy among existing solutions given the same memory budget of switches.

\end{abstract}

%\category{C.2.2}{Computer Communication Networks}{%\hspace*{1.2cm}
%Network Protocols}[Routing Protocols] %\terms{Algorithms, Design,Performance, Reliability}

%\keywords{K1,K2}

\newtheorem{mydef}{theorem}
\section{Introduction}
%(words words words and words. words another word words.

Network traffic monitoring supports fundamental network management tasks,
such as user application identification \cite{pan2012tracking}, anomaly detection \cite{AggregationYinZhang},
%network topology discovering \cite{marchetta2011topology}, 
forensic analysis \cite{xie2005worm},  and traffic engineering \cite{benson2011microte}. Recent studies \cite{csamp} \cite{cSamp-T} \cite{Decor} \cite{AggregationYinZhang}  have addressed two essential requirements of traffic monitoring, namely per-flow monitoring and load distribution.
%These tasks usually require the monitoring infrastructures to provide a complete, reliable and accurate measurement of traffic metrics covering network-wide flows.

Existing traffic measurement tools, e.g. Netflow \cite{Netflow} and sFlow \cite{sflow},  support generic  measurement tasks based on packet sampling, where packets are selected with a given probability.  
%which is based on packet sampling. Each packet is selected with a given probability and the sampled packets are classified  into flows by their 5-tuple , i.e.$\langle$SrcIp, DstIp, SrcPort, DstPort, Protocol$\rangle$. These packet-based  monitoring approaches are reliable enough for applications that requires only rough statistical information.
However, many applications require \emph{per-flow monitoring}, i.e., different monitoring actions are performed on different flows. For example, a monitor may need to examine detailed traces from subsets
of flows \cite{ramachandran2008fast}. Anomaly detection prefers different sampling rates to flow groups \cite{AggregationYinZhang}. A straightforward solution is to let switches store a monitoring rule for each flow. A monitoring rule includes  matching fields and an action applied to the flow, such as sampling with a particular rate or counting packets. As demonstrated in \cite{AggregationYinZhang}, monitoring rule storage consumes non-trivial memory space (tens of thousands entries with aggregation in \cite{AggregationYinZhang}) on a switch. As discussed in many studies \cite{buffalo} \cite{DIFANE} \cite{palette} \cite{payless}, fast switch memory is expensive, power-hungry, and very limited. Therefore rule-based per-flow monitoring has space scalability problem. 

In most networks, a number of routers/switches independently monitor flows. These switches may consume tremendous resources (CPU, memory, bandwidth, etc) to perform monitoring tasks. On the other hand, some flows may not be covered by these switches \cite{csamp}. To resolve this problem, \emph{distributed and collaborative monitoring} \cite{csamp} \cite{cSamp-T} \cite{AggregationYinZhang} has been proposed to allow all switches in the network collaboratively share monitoring load. 

\begin{figure*}[t]
\center

\begin{subfigure}{0.33\textwidth}
\includegraphics[width=\linewidth]{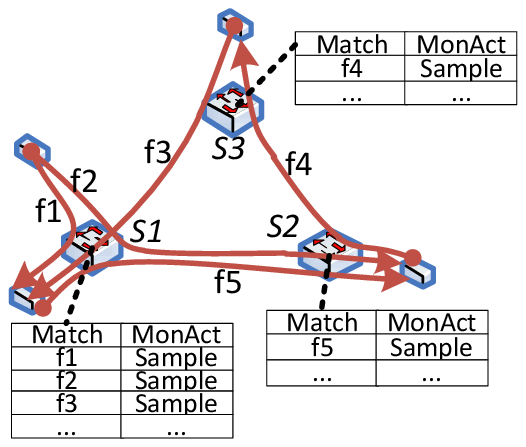}
\vspace{-2.5ex}
\caption{\textbf{Rule-based:} requires large rule storage cost\vspace{2.3em} }
\label{fig:ruleexample}
\end{subfigure}
\begin{subfigure}{0.33\textwidth}
\includegraphics[width=\linewidth]{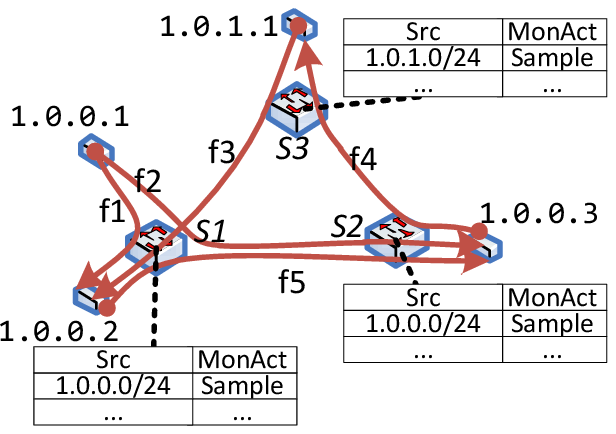}
\vspace{-2.5ex}
\caption{\textbf{Aggregation-based:} flow $f2$ is monitored twice by $S1$ and $S2$. \vspace{1.15em} }
\label{fig:aggexample}
\end{subfigure}
%\begin{subfigure}{0.24\textwidth}
%\includegraphics[width=\linewidth]{gra/sgcSamp.pdf}
%\caption{\textbf{Hash-based}\\Use a global hash function to map packets to $[0,1]$.\\Incapable of
%per-flow monitoring.}
%\label{fig:cSampexample}
%\end{subfigure}
\begin{subfigure}{0.33\textwidth}
\includegraphics[width=\linewidth]{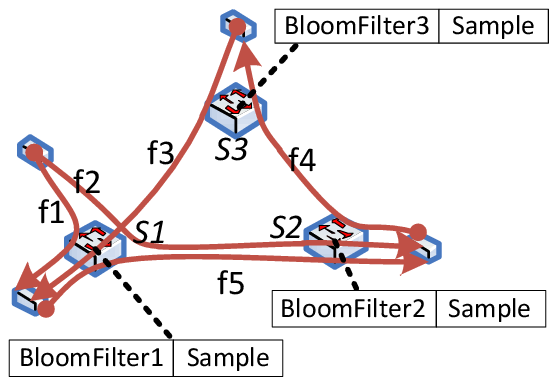}
\vspace{-4ex}
\caption{\textbf{DCM:} $f1$, $f2$, $f5$ match Bloom filter $BF1$;   $f4$ matches $BF2$; $f3$ matches $BF3$.}
\label{fig:BFexample}
\end{subfigure}
\centering
\vspace{-2ex}
\caption{Three distributed and collaborative monitoring methods%\\
\vspace{-1ex}
%\textbf{ (a) Rule-based}: record every rule entry and react according to such rules.\ \ \
%\textbf{ (b) Rule aggregation }: Reduce the number of entries by aggregation,
% simultaneously cause duplicate monitoring. (flow f2 with source IP is $1.0.0.1$ is
% monitored twice at switch $S_1$ and $S_2$.) \ \ \ \
%\textbf{ (c) Hash based}: use a global hash function to map all the packets to a real
%number between $0$ and $1$. A switch only monitors packets whose hash value falls in the
%given range. Incapable of per-flow sampling. \ \ \ \ \
%\textbf{ (d) DCM}: use BloomFilters to enhance memory efficiency.
}
\end{figure*}

Current traffic monitoring tools either are hard to deploy in practical networks or cannot meet both of the two requirements. For example, cSamp \cite{csamp} uses the hash value of the 5-tuple of a packet to distribute sampling load among routers. However, cSamp requires all packets to carry their ingress-egress pairs, which are not available in practical networks \cite{cSamp-T}. The only two approaches that can achieve per-flow monitoring and load distribution are rule-based and aggregation-based flow monitoring.
Figure \ref{fig:ruleexample} shows an example of rule-based monitoring. According to the rules stored on switches, the five flows $f_1$ to $f_5$ will be sampled separately on $S_1$, $S_2$, and $S_3$. As discussed, rule-based monitoring is limited by the switch memory space. Figure \ref{fig:aggexample} shows a solution by aggregate-based approach to sample the five flows. The sampling task of $f_1$, $f_2$, and $f_5$ are assigned to $S_1$, $f_4$ is assigned to $S_2$, and $f_3$ is assigned to $S_3$. Source aggregation saves the memory to store rules. However aggregation still requires a large rule table \cite{AggregationYinZhang}. In addition, potential duplicate samples may occur. For example, $f_5$ is sampled twice on both $S_1$ and $S_2$.

% But studies \cite{ramachandran2008fast,revisiting} have demonstrated the insufficiency of packet sampling for applications that require  precise measurements. Other than packet-based measurements, per-flow monitoring preserves the fidelity of measurement accuracy by classifying packets into flows and perform per-flow monitoring actions.

%Monitoring applications always need excessive resources (e.g. to store monitoring action rules for each entry costs memory, see Fig \ref{fig:ruleexample}). Such requirements usually exceed router capabilities and limits processing rates. To address the resource problems, recent proposals study distributed solutions \cite{csamp,AggregationYinZhang,raza2012measurouting} rather than centralized approaches. Though, the resource on each switch is still limited.

%A possible approach to reduce memory consumption is to to aggregate these per-flow rules, so that the total number of entries in the table can be reduced. Ying Zhang \cite{AggregationYinZhang} aggregates rule entries by source IP prefixes. However, as multiple flows may have identical prefix with interested flows, such aggregation may lead to duplicate monitoring (Fig \ref{fig:aggexample}).

%Coordinated Sampling (cSamp) \cite{csamp} provides higher flow coverage rage ,achieves fine-grained
%network-wide flow coverage goals and efficiently leverages available monitoring capacity on routers.
%cSamp classifies packets by \textit{Origin-Destination (OD) }pairs. The workload of
%monitoring network traffic with a same OD-pair is split into fractions and distributed among switches
%on the path. Each of these fraction is defined by a particular range of hash values of the packet headers.
%A switch only monitors packets whose hash values falls inside the corresponding range. %(Fig \ref{fig:cSampexample}).
%However, cSamp requires each of the router immediately determine the Origin-Destination (OD) pair for each packet
%it sees \cite{cSamp-T}. Thus, cSamp requires upgrades to border routers and modifications to packet headers,
%which hinders cSamp from deployment in ISPs. Meanwhile, cSamp fails to perform per-flow monitoring.

In this paper, we propose \emph{a memory-efficient system for Distributed and Collaborative per-flow Monitoring}, called DCM. DCM uses Bloom filters \cite{bloomfilter} to represent monitoring rules using a small size of memory. It utilizes the tremendous convenience brought by the software defined networking (SDN) paradigm to install a customized and dynamic monitoring tool into the switch data plane. The novel monitoring tool used by DCM is called  \emph{two-stage Bloom filters}, including an admission Bloom filter to accept all flows assigned to the switch and a group of action Bloom filters to perform different measurement actions. SDN also allows DCM to perform updates or reconstruction of the two-stage Bloom filters in the switch data plane. Figure \ref{fig:BFexample} shows an example to use DCM to sampling the five flows.  Switch $S_1$ finds that $f_1$, $f_2$, and $f_5$ match its Bloom filter $BF_1$ and then samples packets of the three flows. Similarly $S_2$ samples $f_4$ and $S_3$ samples $f_3$. Although Bloom filters may introduce false positives, the design of two-stage Bloom filters can reduce the false positive rate to a negligible value with small memory cost.  %Experimental results show that DCM requires
In addition, the SDN controller can detect all false positives and limit their negative influence, due to its central view of the switches and flows. 

The rest of the paper is organized as follows. Section \ref{sec:background} introduces background knowledge of this work. Section \ref{sec:design} presents the system design of DCM. In Section \ref{sec:case} we study how DCM performs two  representative monitoring tasks, namely flow size counting and packet sampling. We also evaluate the performance of DCM for the two tasks using real data center and ISP traffic data and network topologies in the same section. Finally we conclude this work and present future work in Section \ref{sec:future}.

%\textbf{Approach and Intuition}
%Inspired by Bloom Filter, the space-efficient probabilistic data structure, we design a new \textit{Distributed Collaborative Monitoring (DCM)} architecture for Software Defined Networks.

%A key question is how to fairly distribute the flows among switches to allow the witches work collaboratively. Benefiting from the nature of SDN, such allocation  can be computed by the SDN controller .

%For a particular flow, we select one switch on its path and monitor each of its packets on this switch. The number of the flows that should be monitored by a switch is only a little fraction of the total number of flows that it sees.

%The monitoring process for an incoming packet on a switch has two phases:
%(1) \textbf{\textit{Admission}} phase , in which the switch extract the flow information from the packet and find out whether it should be monitored by this switch;
%The \textit{Admission} phase is used to pick out these flows.
%(2) \textbf{\textit{Action}} phase , in which the switch choose the corresponding procedure to process the packet.

\section{Background}
\label{sec:background}
\subsection{Bloom Filter}
A Bloom filter \cite{bloomfilter} $B$ is a simple but space-efficient probabilistic data structure
representing a set of items $S$ and support membership queries.
An item $i$ may match $B$ or fail to match $B$, depends on whether $i$ is in $S$. One key problem of Bloom filters is false positives. The false positive probability of $B$ is $(1-e^\frac{kn}{m})^k$, where $n$ is the size of $S$, $m$ is $B$'s length in bits, and $k$ is the number of hash functions. Bloom filter and its variations have been widely used in the network community to solve various problems, such as distributed caching  \cite{summarycache}, P2P data management \cite{p2p}, unicast and multicast routing \cite{buffalo} \cite{MBF}, and network measurement \cite{blue} \cite{SPIE}.

\subsection{Related works}
Traffic monitoring and measurement support many network management tasks.
The de-facto traffic monitoring standard is packet-based sampling, such as Netflow \cite{Netflow} and sFlow \cite{sflow}. Further, authors in \cite{revisiting} state the importance of using per-flow monitoring. cSamp \cite{csamp} coordinates network-wide routers using hash-based method to
sample packets that carry the OD pair information.  To make cSamp practical, cSamp-T \cite{cSamp-T} removes the assumption of OD pair information on packet headers and Decor \cite{Decor}  applies local information to avoid the use of central controller.
% A omniscient central controller which knows the flow path is needed.

Recently SDN-based traffic monitoring has been studied. OpenSketch \cite{opensketch} is a software defined traffic measurement architecture that applies  sketches for various monitoring tasks. OpenSketch only discusses measurement actions on a single switch. A following paper \cite{moshref2013resource} discusses the tradeoffs between the resource and accuracy of heavy hitter detection.

The SDN data plane scalability problem, i.e., limited rule storage space, has been addressed by recent work. DIFANE \cite{DIFANE} and and Palette \cite{palette} propose to partition or distribute rules over the switches to reduce per-switch rule storage. Payless \cite{payless} and OpenWatch \cite{AggregationYinZhang} use flow aggregates to complete different tasks and reduce the number of rules per-switch.

%  There are a number of other monitor method based on SDN, such as ,   a sketch-based sampling proposal.   The resource of SDN switch is limited. Only limited number of rules can be store    on SDN switch \cite{DIFANE}.    Solutions like \cite{payless} and \cite{AggregationYinZhang} aggregate the rules to save the resource,     \cite{palette} decompose rules into different tables and distribute on different switches.

%Payless's related work is really good and detailed.

%Opensketch\cite{opensketch}->sketch-based sample, programmable measurement, hash-based, software-defined measurement
%cSamp\cite{csamp}-> flow sample, hash-based coordination, OD pairs
%Palette\cite{palette}->distributed table, reduce rule number.
%NetFlow\cite{Netflow}, sFlow\cite{sflow}->sample-based measurement
%Tradeoff\cite{Tradeoff}-> resource, primitives
%openTM\cite{openTM}->distributed query
%daptive Flow Counting\cite{Aggregation}-> aggregation to reduce rule number
%Decor\cite{Decor}->distributed decision making
%Payless\cite{payless}->statistics collection at different aggregation level
%DIFANE\cite{DIFANE}-> reduce the press of central controller
%revisiting\cite{revisiting}-> the importance of network flow monitoring
%cSamp-T\cite{cSamp-T}->without OD pairs.
%ProgME\cite{ProgME}->flowset

\section{System design}
\label{sec:design}
%The goal of Distributed Collaborative Monitoring (DCM)  is to provide a scheme of network measurement, in which the \textit{per-flow} network traffic can be measured \textit{ collaboratively } by multiple routers.

In this section, we detail the design of our Distributed Collaborative Monitoring (DCM) system.

\subsection{Model and Assumptions}
%The goal of DCM is to distribute monitoring duty to switches in a coordinated approach, and perform network-wide flow monitoring. In DCM,

The objective of DCM is to distribute the monitoring duty of the targeted flows to the entire network so that to reduce rule storage and packet processing overhead on switches.  DCM guarantees the following two properties: 1) every packet of a targeted flow should be monitored by at least one switch on its path; 2) if a packet is monitored by more than one switches, duplicate monitoring can be detected.
%In this paper, we say the flow is \textit{admitted} by a switch.

\textbf{System Model}: %The monitoring model is described as below.

$\bullet$ Flows are identified by the 5-tuple, i.e. $\langle$ \emph{SrcIp}, \emph{DstIp}, \emph{SrcPort}, \emph{DstPort}, \emph{Protocol} $\rangle$.

$\bullet$ There is a centralized SDN controller that knows the information (include paths and 5-tuple) of all flows in the network. The controller maintains a monitoring table of the targeted flows and the corresponding monitor actions. Different flows may have different actions. The controller can communicate with a switch to install, update, and delete software-based monitoring tools in the switch data plane.
% $F = \{ f_1,f_2, \cdots f_n \}$.
%For each flow $f$, the controller knows $f$'s path in the network.
%\item The switches shares global hash functions $H_1,H_2,\cdots$. These hash functions map flow 5-tuple to integers and can be calculated by switches in an efficient way.

$\bullet$ A switch installs monitoring tools and processes the packets it encounters. It records measurement results in its local memory and reports the results to the controller periodically. When a switch receives a packet of a new flow, it forwards the flow information to the controller. The controller decides whether,  how, and where the flow should be monitored.
%identify packets by their 5-tuple and perform flow-level action for the packets.
% Common examples are flow size counting \cite{cormode2004improved} , flow sampling , Heavy Hitters \cite{cormode2004improved} and  traffic changes detection \cite{schweller2004reversible}.
%\item The switches maintain memory space to store monitoring rules and results. Switches should be able to maintain these rules and

%$\bullet$ For each monitor action $A$ , Each switch $S$ maintains a set of flows , action set $Act_{S,A}$. Monitor cation $A$ should be processed for $f$ on switch $S$. Such distribution is conducted by the controller intellectually.

%$\bullet$  Each switch $S$ keeps a \textit{admission} set of flows $Adm_{S}$. $Adm_S$ is the union set of all possible $Act_{S,A}$. Only those flows belong to $Adm_S$ need to be processed by a monitor action on $S$.

%$\bullet$ The switches report monitor results to the controller periodically in response to control message.

%\textbf{Assumptions}:
%There are four main assumptions, all of which are reasonable and feasible within current realties of SDN.
We assume that the memory space for monitoring tasks  in a switch is limited while the controller has enough space to store detailed flow information and monitor actions.
%$\bullet$ Switches should be able to maintain and report monitoring results to the controller periodically or in response to control messages.
%$\bullet$ On the controller side, there is abundant memory and computing resource.

%\begin{itemize}
%\item A \textit{monitor action} is a series of procedures which should be performed by a switch, it takes a set of packets as input and produce per-flow monitor results.
%\item For each monitor action $A$, the controller specifies a set of flows $F_A$, $F_A \subset F$. Each packets in these flows
%should be monitored at least once.

%\end{itemize}

%However, due to the fact that Bloom Filter is a probabilistic data structure, false positive admissions may occur. Our evaluation shows that such false positive admissions merely have a insignificant affect on the accuracy of monitoring, which can also be predicted , detected and eliminated by the controller easily.

 \begin{figure}[t]
 \center
 \includegraphics[width=0.9\linewidth]{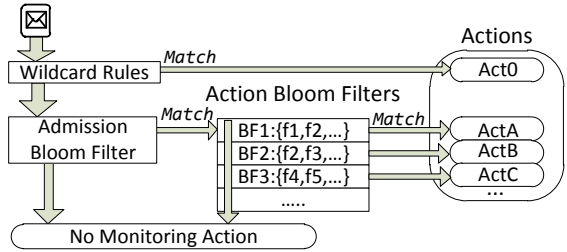}
 \vspace{-1ex}
 \caption{DCM switch data plane}
 \vspace{-2ex}
 \label{fig:DCM}
 \end{figure}

\subsection{DCM Data Plane on Switches}
When a switch receives a packet to forward, the DCM data plane has three steps to process the packet, as shown in Figure \ref{fig:DCM}. The flow-to-filter matching are based on the hash of a 5-tuple.

\textbf{Step 1.} The wild card matching step is to check whether the packet matches one of the wild card monitor rules. A wild card rule applies an action to an aggregate of flows. For example, if DCM wants to sample all flows whose sources are with a same prefix, a wild card can specify such monitoring task in a memory-efficient way. A packet matching a wild card is then be processed by the specified action and skips the remaining steps.
%to deal with rules that may cover large amount of flows. There may be some monitor requirements for a kind of flows that may be easy to identify (i.e, all ICMP packet, all SDN control message, all message from a big aggregation of IP address).

Our main contributions are in the second and third steps using  \emph{two-stage Bloom filtering}.

\textbf{Step 2.} The first part of two-stage Bloom filtering is called the \emph{admission Bloom filter} (admBF). The admBF represents the set of flows which should be monitored but the actions are not specified by any wildcard rule. However, the admBF does not specify any monitoring action. If a packet matches the admBF, it will then be processed to get its action. If a packet does not match the admBF, the DCM data plane knows that it does not belong to any flow under monitoring and then skip the remaining step. Therefore the function of the admBF is to filter the flows that are not of interest.

\textbf{Step 3.} Flows that can match the admBF will be further be checked by the \emph{action Bloom filters} (actBFs) to decide the corresponding monitoring actions. In the example of Figure \ref{fig:DCM}, packets of flows $f_1$ and $f_2$ match $BF_1$ and hence be processed using Action A. Note that a flow may match multiple actBFs. For example, packets of flow $f_2$ match both $BF_1$ and $BF_2$ and have two monitoring actions.

%(4) The action stage , in which a particular monitoring action is performed for the packet.
There are two main purposes to design such two-stage Bloom filtering. First, using the admBF, most packets that are not monitored will be filtered and not checked by the actBFs. Thus it saves the switch processing resource. Second, although some flows which should not be monitored also pass the admBF, the number of flows that are checked by the actBFs significantly reduces. Recall that the false positive probability of a Bloom filter is $(1-e^\frac{kn}{m})^k$ where $n$ is the size of the item set. Two-stage Bloom filtering reduces $n$ for two potential performance gains: 1) give an actBF with size $m$, smaller $n$ will result fewer false positives; 2) give a false positive rate, it needs a actBF with smaller size when $n$ is smaller.

All wildcard rules, admBF, and actBFs are determined by the controller and installed on switches.
Note the DCM component does not perform any packet forwarding task.

%On seeing a packet, a DCM switch first checks whether it is in the wild card rule tables and performs the wild card monitoring actions if necessary. Then it checks whether it belongs to $Adm_S$ by looking up the admission bloom filter, and then find those $Act_{S,A}$s that this flow belongs to by looking up the action bloom filters. Then the corresponding monitor action is performed.

\subsection{Controller Operations}
The DCM component on the controller is responsible for allocation of monitoring load to switches,  Bloom filter construction and updates,  and false positive detection. %The controller conducts such allocation and build bloom filters to store the allocation results. Such bloom filters are sent to the switches periodically.

\subsubsection{Monitoring load allocation}

Given a set of flows to be monitored, the DCM controller distributes the monitoring load to all switches in the network. Such load distribution provide two main advantages. First, compared to today's approach that a switch independently monitor its flows, the collaborative monitoring reduces per-switch computing and recording overhead. Second, the collaborative monitoring may achieve more accurate measurement results. It is because many measurement tools such as Bloom filters and sketches \cite{opensketch} have higher accuracy with lower load.

%The accuracy of many network monitoring applications have strong connections with the measurement load. For example, a Count-Min sketch \cite{cormode2004improved} will demonstrate lower accuracy  when it processes too many different flows. By distributing the measurement load we can achieve higher accuracy. Moreover, the computing capability on a switch also limits the number of different flows that it processes.

The main considerations to design monitoring load distribution can be presented as follows.  When there is a small number of flows to be monitored by an action $A$, we prefer to restrict the monitoring load of $A$ on a few switches rather than all available ones in the network. It is because any switch performing $A$ should store an individual actBF. When many flows need to be monitored by $A$, DCM introduces more switches to balance the load.

For a monitor action $A$, we define a threshold as $\rho_A$. If the number of flows that are processed by $A$ on a switch exceeds $\rho_A$, we consider the switch is overloaded of $A$. Actions may have different threshold because they consume different levels of resources. For example, packet sampling requires more storage space than counting.

%When there is only a small number of flows need to be monitored, most of the $Act_{S,A}$ sets will have a relatively small size. The monitoring work can be performed with capacities of some of the switches, other than all available switches. Thus, we mark a part of the switches as \textit{available} and try to allocate monitor within these switches.
For a new flow $f$ to be monitored by action $A$, if
there is at least one switch on $f$'s path whose current monitoring load of $A$ is less than $\rho_A$, it will be assigned to one of these switches.
Otherwise, the controller assigns $f$ to a switch on $f$'s path that has no acfBF of $A$. In some extreme cases, all  switches on $f$'s path are overloaded, the controller will pick the one with the
minimal load.

%all available switches on $f$'s path are overloaded. In overloaded circumstances, a  new switch is marked as available, and the corresponding memory of $Act_{S,A}$ is then allocated.

%After selecting a switch $S$ to perform monitor action $A$ for flow $f$, $f$ is added to set $Act_{S,A}$ and $Adm_S$.

%We provide an algorithm described as \textbf{BalanceAllocate}. It assigns flows to switches, it balances the work load among switches and using a minimum number of different switches. We use Bloom Filters to maintain these admission sets $Adm_S$ and action sets $Act_{S,A}$,  Namely $BF\text{-}Adm_S$ and $BF\text{-}Act_{S,A}$. Such allocation algorithm is performed on the controller.

Note all allocation results are only stored on the controller. The controller does not communicate with switches at this stage.

\iffalse
\begin{codebox}
\Procname{\proc{ \textbf{BalanceAllocate($BF$-$Decision$)}}}
\li Clear all $BF\text{-}Act$ and $BF\text{-}Adm$
\li Mark all $S$ as unavailable for all action $A$.
\li \While there is a new flow $f_i$ to be monitored with action $A$.
\li \Do   \If All switch on $f_i$'s path are \\ \ \ \ \ \ \ \ \ \ \ \ \ unavailable or overloaded for action $A$
\li \Then choose switch $S_\beta$ with most available resource.
\li Allocate space for $Act_{S_\beta,A}$,
\li  mark $S_\beta$ as available for $A$.
\End
\li Select switch $S_\alpha$ with minimal $Act_{S,A}$ on $f_i$'s path.
\li Add $f_i$ to $Act_{S\alpha,A}$ and $Adm_{S\alpha}$
              \End
\end{codebox}
\fi

\subsubsection{Bloom filter construction and updates}

%\subsubsection{Send and Update Bloom Filters}
Based flows assigned to a switch, the controller computes the admBF and actBFs for different actions of the switch. The false positive rates are pre-determined by the trade-offs between memory cost and accuracy. We recommend that an admBF should be constructed with a very low false positive rate because of two reasons: 1) its false positives may be propagated to actBFs; 2) spending more memory on an admBF is cost-efficient as there is only one admBF on a switch.
After constructing the admBF and actBFs for all switches and actions. The controller encapsulates the
Bloom filters in control messages and sends them to the switches.

The controller also needs to update Bloom filters according to flow dynamics. New flows may join the network and existing flows may end. In addition if the number of flows supported by a Bloom filter increases and the false positive rate is higher than the accuracy requirement, the Bloom filter needs to be reconstructed. It is known that a Bloom filter is easy to perform item addition operations but hard to perform deletion operations. Based on this property, the controller applies a policy called ``\emph{real-time addition and periodical reconstruction}'' (RAPR). When the controller receives a flow to monitor, it will immediately notify the responsible switch to revise its Bloom filters to monitor the new flow. When the controller realizes a flow finishes, it does not perform any operation. In stead, for every period of time $T$, the controller reconstructs all Bloom filters on a switch to remove finished flows and to adjust the filter sizes to meet the accuracy requirement. RAPR guarantees that all flows to monitor will be immediately monitored and reduces the computing and communication cost due to frequency Bloom filter reconstructions. To maintain low false positive rates, the controller also periodically checks each filter using a timeout $T' < T$. If the false positive rate of a filter is higher than its requirement, the controller is also triggered to reconstruct a new filter.

%The monitor requirement may change dynamically, which may cause modification of these bloom filters.
%These change operations can be classified into two categories: (1) Mark a flow $f$ as it should be monitored on switch $S$ by monitor
%action $A$. (2) Mark a flow $f$ that is currently monitored on $S$ by action $A$ as not further necessary.

%Operations in category (1) may be applied easily by simpling insert $f$ into $Adm_S$ and $Act_{S,A}$.
%Operations in category (2) requires removing items from a bloom filter.However, removing an item from a bloom
%filter can not be operated by simply marking the corresponding bits, which may violate other items.
%Meanwhile, the controller is able to detect and eliminate.(e.g, for a
%sampling action, the controller may drop the results that are not necessary.)
% An alternative method for the controller is to record these removing operations and re-construct such bloom filters periodically.
% The controller may keep a table of the removed entries in assistant of removing unwanted parts from the current monitor reports.

\subsubsection{False positive detection}
\label{sec:detection}
Though DCM can control false positive rates, it does not completely eliminate false positives. Thus a flow may be monitored at multiple times on different switches, resulting duplicate measurements.  However, the controller is able to \emph{detect all false positives and limits the negative influence of them}.
%Due to the fact that Bloom Filter is a probabilistic data structure,  false positive admissions and unnecessary monitoring may occur.
%Fortunately,
The controller can maintain copies of Bloom filters installed on switches and the record of flow information. By testing a flow $f$ using all Bloom filters on the switches along the flow path of $f$, the controller may identify all possible duplicate measurements. For example, if $f$ is assigned to be sampled on a switch $s_1$ but  also accidentally matches the Bloom filters on another switch $s_2$ on $f$'s path, the controller can simply knows such false positive and drops all samples of $f$ reported by $s_2$.

%that it calculated for the switches and use it for false positive detection. When controller computes the allocation for flows, it may check and detect potential false positive simply by testing the same flow information (5-tuple) on all the switches on its path. Such detection results can be stored to prepare for correction of correction for monitor results.

\subsection{Discussion of implementation}
The DCM data plane on switches includes three functional components: hash functions, wildcard rule lookup, and Bloom filters. We find that all three components have already been implemented by existing work \cite{BFTCAM, opensketch, buffalo}. In particular, Yu \emph{et al.} \cite{opensketch} uses NetFPGA to implement wildcard lookup and up to 8 hash functions, which are enough to implement the DCM data plane because all actBFs can use a same set of hash functions. The hash function implementation in  \cite{opensketch} is efficient and has no effect on data plane throughput.
Bloom filters can be implemented  either in TCAM \cite{BFTCAM} or in SRAM \cite{buffalo} with slower speed.

% has a very simple architecture. DCM switch data plane can be easily implemented with commodity switch components. OpenSketch \cite{opensketch} provides a complement set of components like hash functions and bloom filters.
%The main component in DCM, Bloom Filter, can be implemented in either SRAM or TCAM. Leveraging among memory resource, processing speed and accuracy, we will be able to implement DCM in an efficient way.

 \begin{figure}[t!]
 \center
 \includegraphics[width=0.85\linewidth]{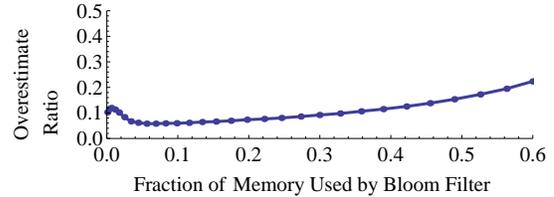}
 \vspace{-1ex}
 \caption{Overestimate ratio v.s. fraction of memory for Bloom filter}
 \vspace{-3ex}
 \label{fig:ratio}
 \end{figure}

\begin{figure*}[t!]
\centering
\begin{subfigure}[b]{0.3\textwidth}
\includegraphics[width=\linewidth]{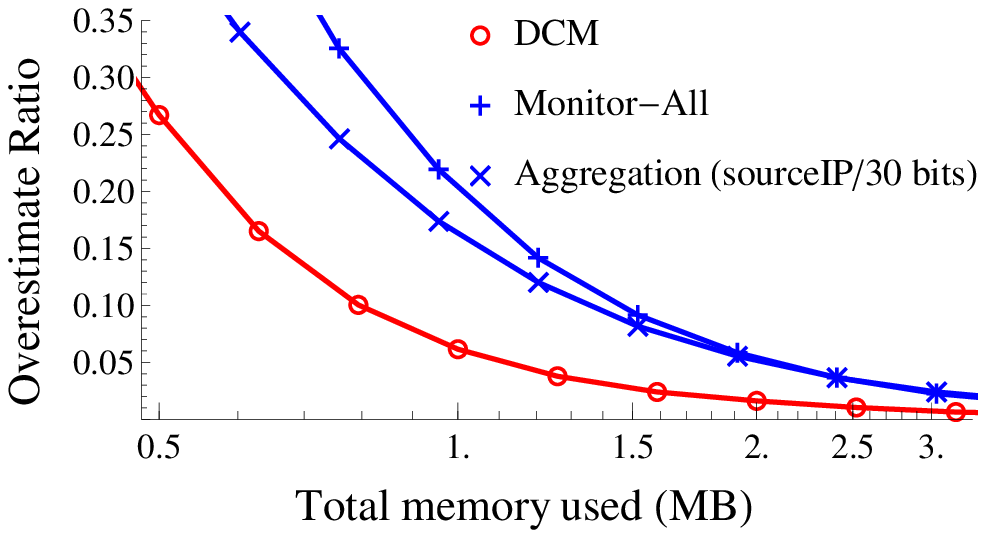}
\vspace{-3ex}
\caption{EDU1}
\label{fig:CntDcCaida}
\end{subfigure}
\begin{subfigure}[b]{0.3\textwidth}
\includegraphics[width=\linewidth]{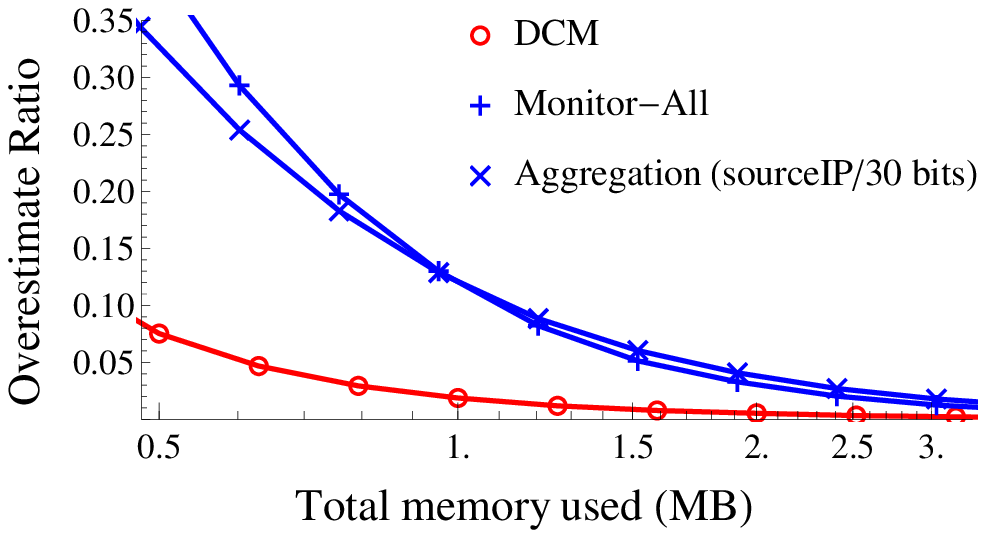}
\vspace{-3ex}
\caption{Fat Tree}
\label{fig:CntDcFT}
\end{subfigure}
\begin{subfigure}[b]{0.3\textwidth}
\includegraphics[width=\linewidth]{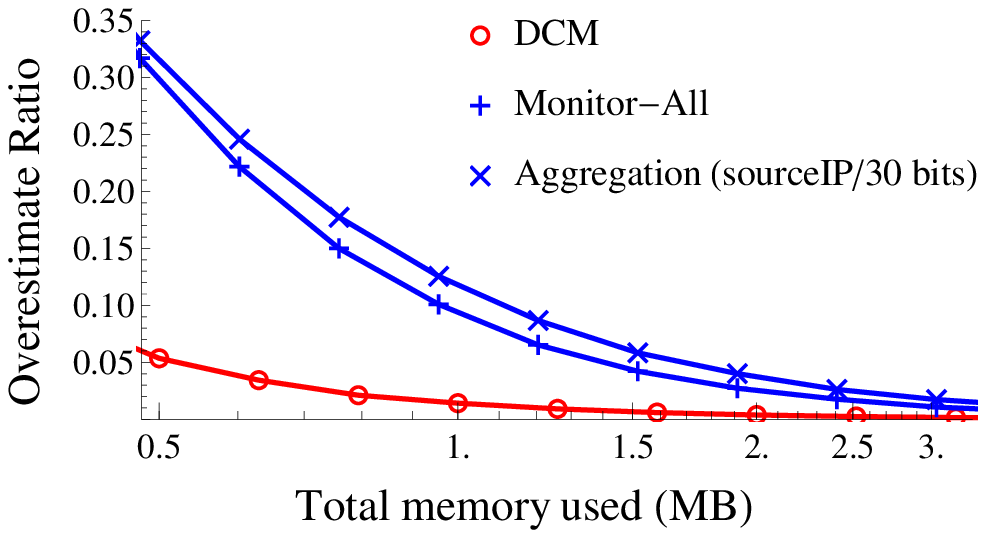}
\vspace{-3ex}
\caption{Rocketfuel}
\label{fig:CntDcRo}
\end{subfigure}
\vspace{-2ex}
\caption{Flow size count: overestimate ratio v.s. total memory consumption}
\vspace{-2ex}
\label{fig:flowSizeCount}
\end{figure*}

\begin{figure*}[t!]
\centering
\begin{subfigure}[b]{0.3\textwidth}
\includegraphics[width=\linewidth]{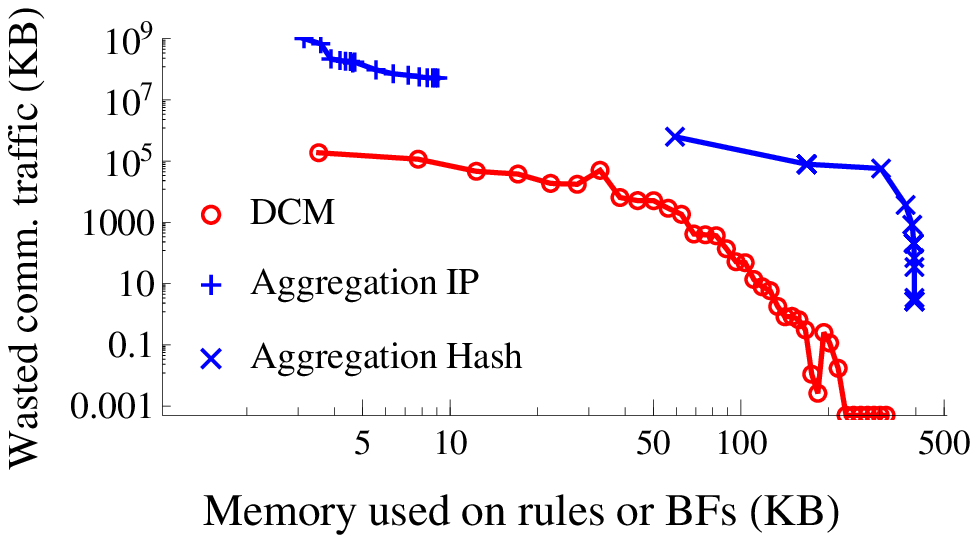}
\vspace{-3ex}
\caption{EDU1}
\label{fig:CntDcCaida}
\end{subfigure}
\begin{subfigure}[b]{0.3\textwidth}
\includegraphics[width=\linewidth]{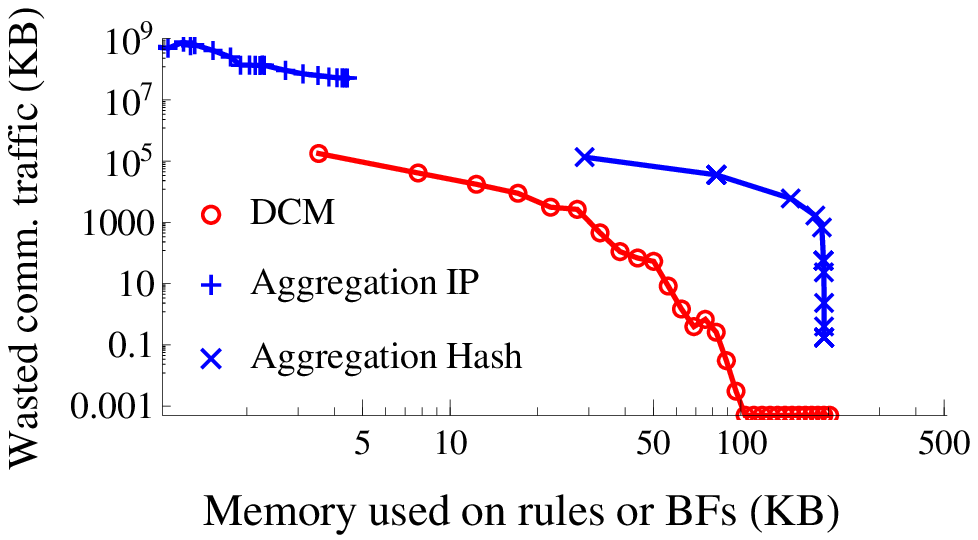}
\vspace{-3ex}
\caption{Fat Tree}
\label{fig:CntDcFT}
\end{subfigure}
\begin{subfigure}[b]{0.3\textwidth}
\includegraphics[width=\linewidth]{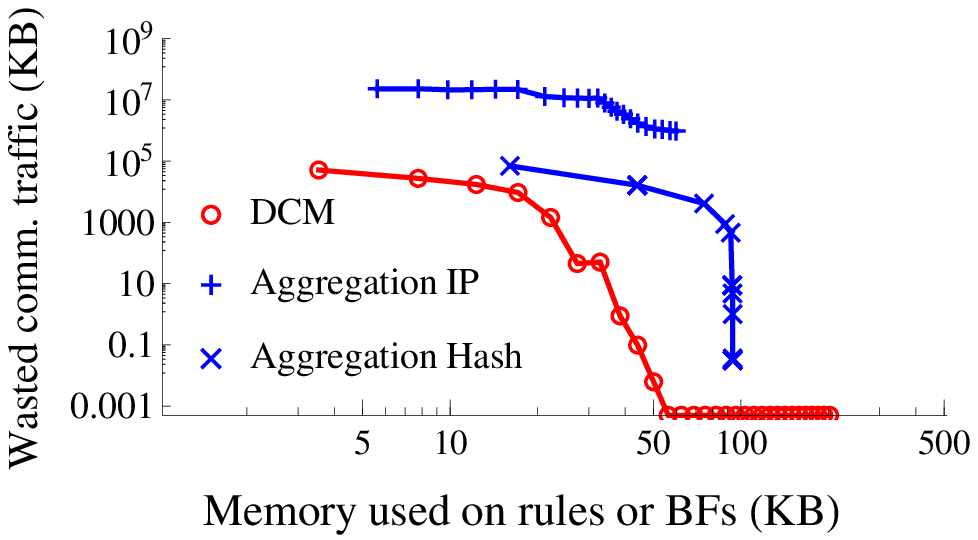}
\vspace{-3ex}
\caption{Rocketfuel}
\label{fig:CntDcRo}
\end{subfigure}
\vspace{-2ex}
\caption{Single-rate sampling: wasted communication traffic v.s. memory used on rules or Bloom filters}
\vspace{-2ex}
\label{fig:singleRateSampling}
\end{figure*}

\section{Case Study and Evaluation}
\label{sec:case}
In this section, we show how DCM supports single-action and multi-action monitoring by studying two representative measurement tasks: flow size counting with Count-Min (CM) sketch and packet sampling.

We also compare DCM with two existing  monitoring methods:
Aggregation-based  monitoring \cite{AggregationYinZhang} and Monitor-All, where Monitor-All is a naive solution that each switch independently monitors all flows. For Monitor-All, we reuse the code of OpenSketch implementation \cite{opensketch}. Rule-based monitoring is not feasible using the memory allocated in all experiments.

We conduct the experiments using two real traffic traces: the EDU1 data from a university data center network \cite{benson2010network} and the CAIDA Anonymized Internet Traces 2013 dataset \cite{caida}. Three network topologies are used: 1)  EDU1, a dual-core, star-shaped topology of the campus data center network in \cite{benson2010network}; 2) Fat-Tree, a typical multi-rooted tree topology \cite{Al-Fares:2008:SCD:1402958.1402967}; and 3) RocketFuel 3967, the router-level ISP network topology of AS 3967 \cite{spring2002measuring}. We apply the EDU1 data on topologies EDU1 and Fat-Tree, and the CAIDA data on RocketFuel.

% conduct the evaluation on two real data center network topologies: (1) UNV1, a dual-core, star-shaped topology provided by \cite{benson2010network}. (2) Fat-Tree \cite{Al-Fares:2008:SCD:1402958.1402967}; and one ISP topology : (3) RocketFuel 3967 \cite{spring2002measuring}. On data center topologies, we use packet traces provided by \cite{benson2010network}, which is observed on UNV1. On ISP network topologies, we use the CAIDA Anonymized Internet Traces 2013 dataset \cite{caida}.

\subsection{Flow Size Counting with Count-Min Sketch}
Flow size counting using the CM sketch \cite{cormode2004improved} has been implemented by OpenSketch \cite{opensketch} for a single switch. Here we discuss how to use DCM for distributed and collaborative monitoring across the network.
%In this case, counting the total length of packets of a flow  is the only monitor action. Thus, action bloom filters are not necessary and the specification stage in DCM data plane is omitted.

A CM sketch is an efficient and probabilistic data structure to support cardinality queries of multiple sets. A CM sketch consists of $k$ arrays $A_1,A_2,\cdots A_k$. An array includes multiple counters. On processing a packet of flow $f$,
 the switch computes $k$ hash values and increments the counter at $A_i[h_i(f)]$. To answer the query for the number of packets of $f$, the  value $MIN\{A_i[h_i(f)]\}$ is returned as
 %each of which is mapped to a counter of each array. : $h_1(f)$ , $h_2(f)$  , $\cdots$  $h_k(f)$. It uses these hash values as index and adds the size of packet into one element in each of the $k$ arrays: $A_1[h_1(f)]$ ,$A_2[h_2(f)]$,$\cdots$ $A_k[h_k(f)]$. The value of $\min$ $\{A_t[h_t(f)]\}$
an estimation of $f$'s size.  CM sketches introduce overestimation. The  accuracy  degrades with the increasing of overall packet numbers and improves with the increasing of memory size to store the sketch.

%stems from the potential conflict between the hash values of $f$ and other flows. The accuracy of this estimation varies against the length and number of such arrays. Count-Min sketch summarizes more precisely with larger memory space.

Flow size counting is a single-action monitoring task. Hence we only need one Bloom filter if no other task is performed at the same time.
In DCM data plane of a switch, a fixed size of memory may be allocated for the Bloom filter and CM sketch. Note that the memory sizes for both the Bloom filter and CM sketch have impact to the accuracy of flow size counting.

We conduct the experiments using the EDU1 data and topology.
Fig \ref{fig:ratio} shows how the average overestimate ratio of the network
changes against the fraction of memory used by the Bloom filter, with total memory limited to 1 MB per switch. We find that the Bloom filter only requires a small fraction of memory (less than 5\%) to achieve the lowest overestimate ratio. When it takes more memory, the accuracy becomes worse because the CM sketch has less memory.

%As Fig \ref{fig:ratio} shows, the overestimate ratio varies while the memory used by bloom filter changes. A very small fraction ( less than 5 \% ) of memory can provide the best performance. This indicates that DCM utilizes memory efficiently.

We compare DCM with  Monitor-All and source IP aggregation using 30-bit mask length in  Fig \ref{fig:flowSizeCount}.
%. and for flow size count with two other approaches. (1) Monitor-All : all packets are monitored multiple times on every switch that it goes through. (2) Aggregation sampling: flows rules are aggregated by source IP prefixes and aggregations are assigned to switches.
We find for all three networks, when provided with same amount of memory, DCM achieves much smaller overestimate ratio than both Aggregation and Monitor-All. Given 2 MB memory,  DCM has very little overestimate. Note that Monitor-All can use all memory for the CM sketch, but its main problem is each switch is responsible for all flows. With more packets mapped to a CM sketch, its  accuracy degrades.

%Meanwhile, to reach the monitor accuracy, DCM requires much less memory than Aggregation and Monitor-All do.

\subsection{Flow Sampling}

\textbf{Single-rate sampling.}
The objective of single-rate sampling is to obtain a fraction of packets from particular flows. As another single-action monitoring, a switch requires only one Bloom filter to identify the flows to monitor.

%In single-rate case, only one monitor action is necessary because all these flows shares the same sample rate.

We compare DCM with two types of aggregation-based methods, for single-rate sampling. Aggregation IP is to group IP addresses by both  source and destination  masks in a certain length.
Aggregation Hash is to aggregate flows by their prefixes of the hash values of 5-tuples. Both DCM and Aggregation have false positives which can be detected. However  communication cost of the report messages from switches to the controller is wasted for the false positive samples.

Figure \ref{fig:singleRateSampling} shows the wasted communication cost versus the memory used on rule or Bloom filter storage. Given the same memory size, DCM causes much less wasted communication cost than Aggregation methods by about  two orders of magnitude. Using 100 KB for Bloom filters, DCM only wastes 100 KB traffic in EDU1 and almost none in Fat-Tree and Rocketfuel.  Aggregation IP can only use a limited range of memory because the mask length cannot be longer than 32. Using 32-bit source and destination masks, false positives still occur due to different port numbers.

%In aggregation solutions, a better granularity may reduce the number of unwanted bits, but such granularity requires larger memory than DCM do.
%Moreover , the granularity is limited by implementation
%reality: the length aggregation is limited by the length of IP address and TCAM keyword length \cite{mishra2010duo}. In our experiments , the granularity is limited to 64 bits.

\begin{figure}[t]
\includegraphics[width=0.85\linewidth]{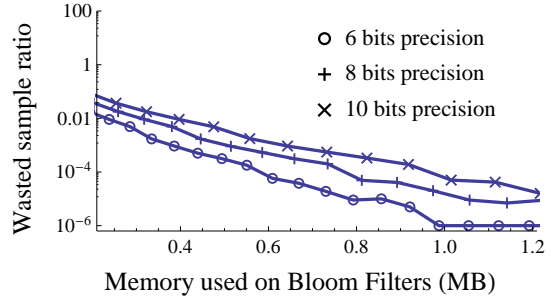}
\vspace{-1ex}
\caption{Multi-rate sampling: wasted sample ratio v.s. memory for Bloom filters }
\label{fig:multiratesampling}
\vspace{-3ex}
\end{figure}

\textbf{Multi-rate sampling.}
As a multi-action monitoring task, multi-rate sampling requires DCM to use multiple actBFs.
Consider a  hash function $H$ maps the packet-related data of packet $p$, e.g., 5-tuple plus sequence number (for TCP) or checksum (Non-TCP), to a value $H(p)$ uniformly in (0, 1).  There are a set of monitor actions $A_1$, $A_2$, $\cdots A_k$, where $A_i$ specifies that $p$ should be sampled if $H(p)$ falls between $\frac{1}{2^{i}}$ and $\frac{1}{2^{i-1}}$. Hence a flow of packets will be sampled by $A_i$ with a rate of $2^{-i}$.
%The hash function $H$ takes the full content of a packet as input and uniformly maps such packet to a real number between 0 and 1.
%$M_i$ is a monitor action that calculates the hash value $H$ of each packet and report those packets with hash
%value falls between $\frac{1}{2^{i}}$ and $\frac{1}{2^{i-1}}$ as sampled. $M_i$ carries uniform sample action
%for flow $f$ with sample rate $2^{-i}$.
%Note that any of these intervals $(\frac{1}{2^i},\frac{1}{2^{i-1}})$  does not intersect, every packet can only be sampled once by one of $M_1,M_2, \cdots,M_k$.
%For a given ratio $p$, we can easily find the binary expression of $p$, and construct a corresponding number sequence ${p_l}$. Let the $b_i$ be the position of the $i$-th $1$ in $p$'s binary expression.
For a given ratio $p$, we construct a number sequence $b_1$,$b_2$,$\cdots$, where $b_t$ is the position of the  $t$-th $1$ in $p$'s binary expression.
Thus, $p = \sum 2^{-b_t}$.
For example , if a flow should be sampled with rate $\frac{11}{16} =
(0.1011)_2$, its 5-tuple can match three actBFs whose actions are $A_1$, $A_3$, and $A_4$.
%, since the first, third and fourth position of the binary expression is 1.
%Assume flow $f$ is assigned with ratio $p$, by carrying out monitor actions $\{ M_{b_1},M_{b_2},\cdots \}$ collaboratively for $f$
%, we are able to get a sample of $f$ with rate $p$.
There is no duplicate sampling by different monitor actions
, because the hash of a particular packet will fall into the interval of at most one action $A_{i}$.
Note that an coefficient can always be applied on a sample action to get lower rate.

In our evaluation, each flow is given a random sample rate. We vary the precision of the rate binary expression  by 6, 8, and 10 bits.
Due to  false positives,  a packet could be sampled on multiple switches. Duplicate samples can always be detected by the controller as discussed in Section \ref{sec:detection}. These duplicates are considered wasted samples.  Figure \ref{fig:multiratesampling} shows the wasted sample ratio versus the memory for Bloom filters. When more than 1 MB is used, multi-rate sampling of all levels of precision has negligible wasted samples.
%that such oversample rate is trifling: when provided with 1MB memory, the average oversample ratio is only $\sim \frac{1}{10^6}$
%for sample rates of 6 bits precision. DCM performs multi-rate sampling precisely.

\section{Conclusion and Future Work}
\label{sec:future}
We have proposed a  Distributed Collaborative Monitoring (DCM) system for SDN-enabled flow monitoring and measurement. We have designed a novel two-stage Bloom filters as the DCM data plane to represent monitoring rules in an efficient and reliable way. Experiments using real traffic data and network topologies show that DCM provides accurate and memory-efficient flow measurement for two representative tasks, i.e., flow size counting and packet sampling.
%Compared to current solutions, DCM provides network-wide flow coverage, achieves fine monitor accuracy. We demonstrated how DCM can be applied to perform monitor tasks like count flow size and multi-rate sampling. Our evaluation shows DCM achieves precise monitor goal with high memory efficiency.

%\section{Discussion and Future Work}
%adjust the length of bloom filters , facing dynamic allocation.
%The false positive rate of DCM highly depends on the configuration of Bloom filters, i.e., lengths and number of hash functions.
%the work load (i.e., number of flows that are supposed to pass the bloom filters).
%Particularly,
In the future, we will explore the following problems.

\textbf{DCM configuration under traffic dynamics.}
%For a given number of flows to monitor and fixed memory budget on each switch, there exists a Bloom filter configuration for fewest false positives.
In practice, monitoring load may change dynamically, which motivates us to design sophisticated DCM data plane construction and update algorithms.
%a special scenario that adjust the bloom filter lengths to reach higher monitor precision. Such procedure involves computing the best bloom filter parameters %in response to the work load and re-construction of existing bloom filters.
We will quantitatively analyze and evaluate the impact of different DCM data plane configurations by varying a number of parameters, including size and number of Bloom filters, fractions of memory allocated for admBf and actBFs, and reconstruction period.
%quantitative analysis ,
%DCM precision varies against several parameters of data plane. The memory resource on a switch should be allocated to one of admission bloom filter, action bloom filter and monitor applications. The distribution of monitor actions may also be further improved by a more intelligent allocation algorithm.

\textbf{Load assignment optimization.}
We also plan to design and analyze different load assignment algorithms to achieve optimal load balance,  memory efficiency, and accuracy.
%answer the following questions conduct a detailed quantitative analysis on these questions in order to reach even higher monitor accuracy and memory efficiency: (1) What is the best ratio to allocate memory among different stages in DCM data plane? (2) How to allocate monitor actions to flows to distribute them network-wide? We believe such optimization will improve the overall monitor efficiency and accuracy of DCM.

\textbf{Prototype implementation.}
We plan to implement a DCM prototype and try to apply it for real traffic monitoring tasks in our campus network, where OpenFlow switches have already been deployed for other network management purposes.
%The implementation of DCM and its deployment in realistic network environment remains to be studied. We will evaluate and analysis the realistic performance and usability of DCM.

% implementation

%In dynamic networking settings, the flow set that people want to monitor
%and measure may change over time. In our current work, we does not discuss
%how to update the Bloom Filters when interested flow set changes.
%Using Counting Bloom Filter\cite{summarycache}, people can insert or
%delete elements in the set incrementally with modest overhead.
%In our future work, we would like to develop a lightweight
%but efficient method to dynamically update the Bloom Filters,
%with the aid of Counting Bloom Filter.

%False positive in Bloom Filter is not desired in our system.

%The accuracy of Bloom Filter could be further improved if false
%positive can be detected. If the central controller detects duplication,
%reconstructing the corresponding Bloom Filter that cause the duplication
%could eliminate this particular false positive and hence improving the accuracy.
% One possible way is cascading a table after the Bloom Filter to
% record the queries that case false positive.
% A flow is sampled only when it passes the Bloom Filter
% and it is not in the following table.
% We plan to test the feasibility of such method and measure the improvement in a quantitative way.

% \subsection{}

%wild card rule is used to process flows that can be easily agggregated.....

%bloom filter length growth.

%.....false positive re-construction.

%memory is good, what about other resources?

%\input{draft}

{\small
\bibliographystyle{abbrv}
\bibliography{pap}}
\balancecolumns
\end{document}